\documentclass[prl,twocolumn]{revtex4-2}
\usepackage{mathrsfs}
\usepackage{amsmath}
\usepackage{amssymb}
\usepackage{graphicx}
\usepackage{esint}
\usepackage{bm}
\usepackage{braket}

\usepackage{hyperref}
\usepackage{color}

\definecolor{myblue}{RGB}{0,0,255}
\hypersetup{
    colorlinks,
    citecolor=myblue,
    linkcolor=myblue,
    urlcolor=myblue
}

\begin{document}
\title{Thermodynamic Uncertainty Relation for General Open Quantum Systems}
\author{Yoshihiko Hasegawa}
\email{hasegawa@biom.t.u-tokyo.ac.jp}

\affiliation{Department of Information and Communication Engineering, Graduate
School of Information Science and Technology, The University of Tokyo,
Tokyo 113-8656, Japan}
\date{\today}
\begin{abstract}
We derive a thermodynamic uncertainty relation for general open quantum
dynamics, described by a joint unitary evolution on a composite system
comprising a system and an environment. By measuring the environmental
state after the system-environment interaction, we bound the counting
observables in the environment by the survival activity, which reduces
to the dynamical activity in classical Markov processes. Remarkably,
the relation derived herein holds for general open quantum systems
with any counting observable and any initial state. Therefore, our
relation is satisfied for classical Markov processes with arbitrary
time-dependent transition rates and initial states. We apply our relation
to continuous measurement and the quantum walk to find that the quantum
nature of the system can enhance the precision. Moreover, we can make
the lower bound arbitrarily small by employing appropriate continuous
measurement. 
\end{abstract}
\maketitle
\emph{Introduction.}---Higher precision demands more resources. Although
this fact is widely accepted, it has only recently been theoretically
proved. The thermodynamic uncertainty relation (TUR) \cite{Barato:2015:UncRel,Gingrich:2016:TUP,Pietzonka:2016:Bound,Horowitz:2017:TUR,Pigolotti:2017:EP,Garrahan:2017:TUR,Dechant:2018:TUR,Barato:2018:PeriodicTUR,Terlizzi:2019:KUR,Hasegawa:2019:CRI,Hasegawa:2019:FTUR,Vu:2019:UTURPRE,Vu:2020:TURProtocolPRE,Dechant:2020:FRIPNAS,Vo:2020:TURCSL}
(see \cite{Horowitz:2019:TURReview} for a review) serves as a theoretical
basis for this notion, and it states that current fluctuations, quantified
by a coefficient of variation, are bounded from below by thermodynamic
costs, such as entropy production and dynamical activity. It predicts
the fundamental limit of biomolecular processes and thermodynamic
engines, and can be applied to infer the entropy production of thermodynamic
systems in the absence of detailed knowledge about them \cite{Seifert:2019:InferenceReview,Li:2019:EPInference,Manikandan:2019:InferEPPRL,Vu:2020:EPInferPRE,Otsubo:2020:EPInferPRE}.

Much progress has been made on the TUR for classical stochastic thermodynamics.
Quantum analogs of the TUR have been recently carried
out, but they are still at an early stage. Many existing studies on
quantum TURs \cite{Erker:2017:QClockTUR,Brandner:2018:Transport,Carollo:2019:QuantumLDP,Liu:2019:QTUR,Guarnieri:2019:QTURPRR,Saryal:2019:TUR,Hasegawa:2020:QTURPRL,Friedman:2020:AtomicTUR}
are concerned with rather limited situations. In the first place,
although an observable of interest in the TUR of classical stochastic
thermodynamics is well defined, there is no consensus regarding specific
observables that should be bounded in the TURs of quantum systems.
In the present Letter, we obtain a TUR for general open quantum systems,
which can be described as a joint unitary evolution of a composite
system comprising a principal system and an environment. Using the
composite representation, we formulate a TUR in open quantum systems
as a bound for the environmental measurement by using the quantum
estimation theory \cite{Helstrom:1976:QuantumEst,Hotta:2004:QEstimation,Paris:2009:QFI,Liu:2019:QFisherReviewJPA}.
The obtained relation exhibits remarkable generality. It
holds for general open quantum dynamics with any counting observable
and any initial density operator. Moreover, our bound applies to
any classical time-dependent Markov process and counting observable.
Our TUR bounds the fluctuations in the counting observables by a quantity
referred to as a survival activity, which reduces to the dynamical
activity \cite{Maes:2020:FrenesyPR} of classical Markov processes
in a particular limit. We apply our TUR to the continuous measurement
and the quantum walk, and find that the system's quantum nature can
enhance the precision of the observables and that an arbitrary small
lower bound of the fluctuations can be achieved by employing appropriate
continuous measurement.

\begin{figure*}
\includegraphics[width=15cm]{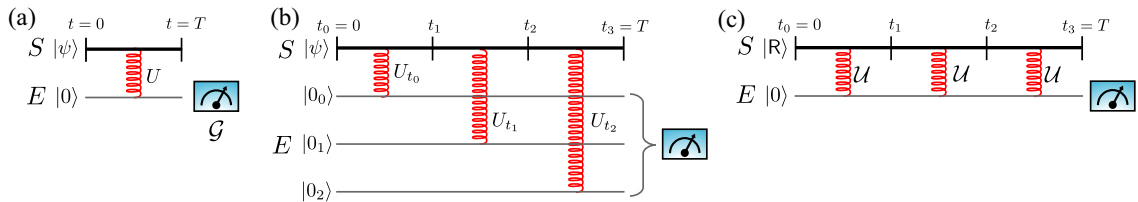} \caption{Principal system $S$ and environment $E$. (a) Basic model. The initial states
of $S$ and $E$ are $\ket{\psi}$ and $\ket{0}$, respectively. The
composite system $S+E$ undergoes a unitary transformation $U$, and
$E$ is measured by the observable $\mathcal{G}$. (b) Continuous
measurement case ($N=3$). The initial states of the principal system
and environment are $\ket{\psi}$ and $\ket{0_{2},0_{1},0_{0}}$,
respectively. The initial sub-state $\ket{0_{k}}$ interacts with
$S$ within the time interval $[t_{k},t_{k+1}]$ via a unitary operator
$U_{t_{k}}$. The measurement record is obtained by measuring $E$
at $t=T$. (c) Quantum walk case. The initial states of chirality
(principal system) and position (environmental system) are $\ket{\mathsf{R}}$
and $\ket{0}$, respectively. The principal and environmental systems
interact at each step via a unitary operator $\mathcal{U}$. The position
is obtained by measuring $E$ at step $t=T$. \label{fig:input_output}}
\end{figure*}

\emph{Results.}---Let us consider a system $S$ and an environment
$E$. The environment comprises an orthonormal basis $\{\ket{m}\}_{m=0}^{M}$.
We assume that the initial states of $S$ and $E$ are $\ket{\psi}$
and $\ket{0}$, respectively. Because $S$ and $E$ interact from
$t=0$ to $t=T$ via a unitary operator $U$ acting on $S+E$, the
state of $S+E$ at $t=T$ is $\ket{\Psi(T)}=U\ket{\psi}\otimes\ket{0}$
(Fig.~\ref{fig:input_output}(a)). Typically, in open quantum systems,
the primary object of interest is the state of the \emph{principal
system} $S$ after the interaction. In contrast, we here focus on
the state of the \emph{environment} $E$ after the interaction. For
example, in continuous monitoring of photon emissions in open quantum
systems, photons emitted into the environment during $[0,T]$ can
be equivalently obtained by measuring the environment at final time
$t=T$. Therefore, the environment includes all information about
the measurement records of the emitted photon.

Suppose that a measurement is performed on the environment at $t=T$
by an Hermitian operator $\mathcal{G}$ (Fig.~\ref{fig:input_output}(a)).
Here, $\mathcal{G}$ admits the eigendecomposition $\mathcal{G}=\sum_{m}g(m)\ket{\phi_{m}}\bra{\phi_{m}}$,
where $\ket{\phi_{m}}$ and $g(m)$ are the eigenvector and eigenvalue
of $\mathcal{G}$, respectively. Using $\ket{\phi_{m}}$, the state
of $S+E$ at $t=T$ can be expressed as \cite{Nielsen:2011:QuantumInfoBook}
\begin{equation}
\ket{\Psi(T)}=U\ket{\psi}\otimes\ket{0}=\sum_{m=0}^{M}V_{m}\ket{\psi}\otimes\ket{\phi_{m}}.\label{eq:Psi_evolution}
\end{equation}
Here, $V_{m}\equiv\braket{\phi_{m}|U|0}$ is the action on $S$ associated
with a transition in $E$ from $\ket{0}$ to $\ket{\phi_{m}}$ and
satisfies $\sum_{m=0}^{M}V_{m}^{\dagger}V_{m}=\mathbb{I}_{S}$, where
$\mathbb{I}_{S}$ is an identity operator in $S$. Although Eq.~\eqref{eq:Psi_evolution}
is a simple interaction model, it can describe the general open quantum
dynamics starting from pure states. When tracing out $E$ in Eq.~\eqref{eq:Psi_evolution},
we obtain the Kraus representation $\rho(T)=\mathrm{Tr}_{E}\left[\ket{\Psi(T)}\bra{\Psi(T)}\right]=\sum_{m=0}^{M}V_{m}\ket{\psi}\bra{\psi}V_{m}^{{\dagger}}$.
We hereafter assume that 
\begin{equation}
g(0)=0,\label{eq:F0_vanish}
\end{equation}
whose physical meaning is explained as follows. For illustrative purposes,
suppose that $\ket{\phi_{m}}=\ket{m}$. Here, $g(0)$ is associated
with $\ket{\phi_{0}}=\ket{0}$ in $E$ after the interaction {[}Eq.~\eqref{eq:Psi_evolution}{]}.
When the state of the environment after the interaction is $\ket{0}$,
the environment remains unchanged before and after the interaction.
For the photon counting problem, $g(m)$ encodes the number of photons
emitted into the environment. In this case, ``no change'' in the
environment corresponds to no photon emission. Therefore, the condition
of Eq.~\eqref{eq:F0_vanish} is naturally satisfied by a photon counting
case. Because the condition of Eq.~\eqref{eq:F0_vanish} constitutes
the minimum assumption for any counting statistics, we refer to observables
satisfying Eq.~\eqref{eq:F0_vanish} as \emph{counting observables}.
For general open quantum dynamics and measurement of the environment,
we wish to find the bound for the fluctuation of $\mathcal{G}$. Let
$\rho$ be the initial density operator of $S$. The mean and variance
of $\mathcal{G}$ are $\braket{\mathcal{G}}\equiv\braket{\Psi(T)|\mathbb{I}_{S}\otimes\mathcal{G}|\Psi(T)}$
and $\mathrm{Var}[\mathcal{G}]\equiv\braket{\mathcal{G}^{2}}-\braket{\mathcal{G}}^{2}$,
respectively. Using the quantum Cram\'er--Rao inequality \cite{Helstrom:1976:QuantumEst,Hotta:2004:QEstimation,Paris:2009:QFI,Liu:2019:QFisherReviewJPA}, we find the following
bound for the coefficient of variation of $\mathcal{G}$: 
\begin{equation}
\frac{\mathrm{Var}\left[\mathcal{G}\right]}{\braket{\mathcal{G}}^{2}}\ge\frac{1}{\Xi},\label{eq:main_result}
\end{equation}
where 
\begin{equation}
\Xi\equiv\mathrm{Tr}_{S}[(V_{0}^{\dagger}V_{0})^{-1}\rho]-1.\label{eq:Xi_def}
\end{equation}
Equation~\eqref{eq:main_result} is the first main result of this
Letter, and its proof is provided in the Derivation section near the
end. Equation~\eqref{eq:main_result} holds for any open quantum
system as long as $V_{0}^{\dagger}V_{0}$ is positive definite, any
counting observable $\mathcal{G}$, and any initial density operator
$\rho$ in $S$. Unless $V_{0}^{\dagger}V_{0}$ is positive definite,
$(V_{0}^{\dagger}V_{0})^{-1}$ is not well defined in Eq.~\eqref{eq:Xi_def},
indicating that $V_{0}$ should be a full-rank matrix \cite{Supp:PhysRev}. \nocite{Streltsov:2017:CoherenceReview,Scully:2011:CoherentHE,Holubec:2018:QHE}
Equation~\eqref{eq:main_result} also holds for any (time-dependent)
classical Markov process with any counting observable. $V_{0}$
is an operator corresponding to ``no change'' in the environment,
and therefore, the expectation of the inverse of $V_{0}^{\dagger}V_{0}$
quantifies activity of the dynamics. For classical Markov processes,
$\Xi$ becomes the reciprocal expectation of the survival probability,
which reduces to the dynamical activity \cite{Maes:2020:FrenesyPR}
in a short time limit {[}see Eq.~\eqref{eq:Xi_classical}{]}. Therefore,
we refer to $\Xi$ as a \emph{survival activity} in the present Letter.
The generality of the bound implies that $\Xi$ is a physically important
quantity. When there are more than one mutually commutable counting
observables $\mathcal{G}_{i}$, we can obtain a multidimensional variant
of Eq.~\eqref{eq:main_result}, as derived in \cite{Dechant:2019:MTUR,Timpanaro:2019:EFTTUR}
(see \cite{Supp:PhysRev} for details).

We note the differences between the present TUR and related quantum
TURs. Reference~\cite{Carollo:2019:QuantumLDP} obtained the TUR
for quantum jump processes. In this case, the TUR was derived using
a semi-classical approach via the large deviation principle for $T\to\infty$.
Reference~\cite{Guarnieri:2019:QTURPRR} used the classical Cram\'er--Rao
inequality to derive a TUR in quantum nonequilibrium steady states.
Their bound concerns instantaneous currents, which are defined by
current operators and derived under a steady-state condition. Recently,
we derived a quantum TUR for arbitrary continuous measurement satisfying
a scaling condition \cite{Hasegawa:2020:QTURPRL}. However, the bound
of Ref.~\cite{Hasegawa:2020:QTURPRL} requires a steady-state condition
under Lindblad dynamics, whereas Eq.~\eqref{eq:main_result} is satisfied
for any open quantum dynamics.

We also comment on the relation between the quantum speed limit (QSL)
\cite{Mandelstam:1945:QSL,Margolus:1998:QSL,Deffner:2017:QSLReview}
and the TUR. The QSL is concerned with the evolution speed, and quantum
estimation theory has been reported to play an important role in the
QSL \cite{Jones:2010:QSL,Frowis:2012:FisherQSL,Taddei:2013:QSL}.
While the QSL focuses on the transformation of the \emph{principal}
system, the TUR in this Letter is concerned with the evolution of
the \emph{environment}. Therefore, the QSL and TUR bound the evolution
of the complementary states by thermodynamic quantities.

\emph{Quantum continuous measurement.}---To observe the physical
meaning of the main result, we apply Eq.~\eqref{eq:main_result}
to continuous measurement in open quantum systems. Let us consider
a Lindblad equation \cite{Lindblad:1976:Generators,Breuer:2002:OpenQuantum}:
\begin{equation}
\frac{d\rho}{dt}=-i\left[H,\rho\right]+\sum_{m=1}^{M}\mathcal{D}(\rho,L_{m}),\label{eq:Lindblad_def}
\end{equation}
where $H$ is a Hamiltonian, $\mathcal{D}(\rho,L)\equiv\left[L\rho L^{\dagger}-\left\{ L^{\dagger}L,\rho\right\} /2\right]$
is a dissipator, and $L_{m}$ ($1\le m\le M$) is an $m$th jump operator
with $M$ being the number of jump operators ($[\bullet,\bullet]$
and $\{\bullet,\bullet\}$ denote the commutator and anticommutator,
respectively). Within a sufficiently small time interval $[t,t+\Delta t]$,
one possible Kraus representation for Eq.~\eqref{eq:Lindblad_def}
is $\rho(t+\Delta t)=\sum_{m=0}^{M}X_{m}\rho(t)X_{m}^{\dagger}$,
where 
\begin{align}
X_{0} & \equiv\mathbb{I}_{S}-i\Delta tH-\frac{1}{2}\Delta t\sum_{m=1}^{M}L_{m}^{\dagger}L_{m},\label{eq:Jump_V0_def}\\
X_{m} & \equiv\sqrt{\Delta t}L_{m}\;\;\;(1\le m\le M).\label{eq:Jump_Vc_def}
\end{align}
$X_{m}$ satisfies the completeness relation $\sum_{m=0}^{M}X_{m}^{\dagger}X_{m}=\mathbb{I}_{S}$
upto $O(\Delta t)$.

By using the input-output formalism \cite{Guta:2011:AsympNorm,Gammelmark:2014:QCRB,Macieszczak:2016:QMetrology}
(see Ref.~\cite{Gross:2018:ContMeasQubit} for comprehensive description),
we can describe the time evolution induced by the Kraus operators
of Eqs.~\eqref{eq:Jump_V0_def} and \eqref{eq:Jump_Vc_def} as an
interaction between the system $S$ and environment $E$. Let $N$
be a sufficiently large natural number. We discretize the time by
dividing the interval $[0,T]$ into $N$ equipartitioned intervals,
and define $\Delta t\equiv T/N$ and $t_{k}\equiv k\Delta t$. We
assume that the environmental orthonormal basis is $\ket{m_{N-1},...,m_{0}}$,
where a subspace $\ket{m_{k}}$ interacts with $S$ within the time
interval $[t_{k},t_{k+1}]$ via a unitary operator $U_{t_{k}}$ (Fig.~\ref{fig:input_output}(b)).
When the initial states of $S$ and $E$ are $\ket{\psi}$ and $\ket{0_{N-1},...,0_{0}}$,
respectively, the state of $S+E$ at time $t=T$ is 
\begin{align}
\ket{\Psi(T)} & =U_{t_{N-1}}\cdots U_{t_{0}}\ket{\psi}\otimes\ket{0_{N-1},...,0_{0}}\nonumber \\
 & =\sum_{\boldsymbol{m}}X_{m_{N-1}}\cdots X_{m_{0}}\ket{\psi}\otimes\ket{m_{N-1},...,m_{0}},\label{eq:Psi_photon_def}
\end{align}
where $\boldsymbol{m}\equiv[m_{N-1},...,m_{0}]$, $X_{m_{k}}\equiv\braket{m_{k}|U|0_{k}}$
is an operator associated with the action of jumping from $\ket{0_{k}}$
to $\ket{m_{k}}$ in $E$, and $\ket{m_{N-1},...,m_{0}}$ provides
the record of jump events. When the environment is measured using
$\ket{m_{N-1},...,m_{0}}$ as a basis, the unnormalized state of the
principal system is $X_{m_{N-1}}\cdots X_{m_{0}}\ket{\psi}$, which
is referred to as a quantum trajectory conditioned on the measurement
record $\boldsymbol{m}=[m_{N-1},...,m_{0}]$. The evolution of a quantum
trajectory is given by a stochastic Schr\"odinger equation \cite{Molmer:1993:MonteCarlo,Wiseman:1996:QJ,Daley:2014:QJReview}:
$d\rho=-i[H,\rho]dt+\sum_{m=1}^{M}\left(\rho\mathrm{Tr}_{S}\left[L_{m}\rho L_{m}^{\dagger}\right]-\frac{\left\{ L_{m}^{\dagger}L_{m},\rho\right\} }{2}\right)dt+\sum_{m=1}^{M}\left(\frac{L_{m}\rho L_{m}^{\dagger}}{\mathrm{Tr}_{S}[L_{m}\rho L_{m}^{\dagger}]}-\rho\right)d\mathcal{N}_{m}$,
where $d\mathcal{N}_{m}$ is a noise increment equal to $1$ when
the $m$th jump event is detected between $t$ and $t+dt$; otherwise,
$d\mathcal{N}_{m}=0$. Given the complete history of $[\mathcal{N}_{m}(t)]_{m=1}^{M}$,
the conditional expectation is $\braket{d\mathcal{N}_{m}(t)}=\mathrm{Tr}_{S}[L_{m}\rho(t)L_{m}^{\dagger}]dt$,
where $\rho(t)$ is a solution of the stochastic Schr\"odinger equation. 

We consider a counting observable $\mathcal{G}$ in the continuous
measurement, which counts the number of jump events within $[0,T]$.
When expressed classically, we may write $\mathcal{G}=\sum_{m=1}^{M}G_{m}\mathcal{N}_{m}$,
where $G_{m}\in\mathbb{R}$ is the weight of the $m$th jump and $\mathcal{N}_{m}=\int_{0}^{T}d\mathcal{N}_{m}$
is the number of $m$th jumps during $[0,T]$. Because the state of
$S+E$ at time $t=T$ is given by Eq.~\eqref{eq:Psi_photon_def},
$\mathcal{G}$ can be defined quantum mechanically by $\mathcal{G}=\sum_{\boldsymbol{m}}g(\boldsymbol{m})\ket{\boldsymbol{m}}\bra{\boldsymbol{m}}$.
Because $\boldsymbol{m}$ is a record of jump events, $g(\boldsymbol{m})$
should be defined so that it counts and weights each jump event according
to the classical definition of $\mathcal{G}$. When the environment
remains unchanged before and after the interaction, $\mathcal{N}_{m}=0$
for all $m\ge1$. Therefore, $\mathcal{G}$ naturally satisfies the
condition of Eq.~\eqref{eq:F0_vanish}. $X_{0}$ in Eq.~\eqref{eq:Jump_V0_def}
corresponds to a no-jump event within $[t,t+\Delta t]$. Because $V_{0}$
in Eq.~\eqref{eq:main_result} corresponds to the action associated
with no jump events within $[0,T]$, it is given by $V_{0}=\lim_{N\to\infty}X_{0}^{N}$.
We obtain $V_{0}=e^{-T\left(iH+\frac{1}{2}\sum_{m=1}^{M}L_{m}^{\dagger}L_{m}\right)}$
and the survival activity is expressed as 
\begin{equation}
\Xi=\mathrm{Tr}_{S}\left[e^{T\left(iH+\frac{1}{2}\sum_{m=1}^{M}L_{m}^{\dagger}L_{m}\right)}e^{T\left(-iH+\frac{1}{2}\sum_{m=1}^{M}L_{m}^{\dagger}L_{m}\right)}\rho\right]-1.\label{eq:Xi_Lindblad_def}
\end{equation}
When $H$ and $L_{m}$ depend on time, $\Xi$ is formally given by
\cite{Supp:PhysRev} 
\begin{align}
\Xi & =\mathrm{Tr}_{S}\Bigl[\overline{\mathbb{T}}e^{\int_{0}^{T}dt\,iH(t)+\sum_{m=1}^{M}L_{m}^{\dagger}(t)L_{m}(t)/2}\nonumber \\
 & \mathbb{T}e^{\int_{0}^{T}dt\,-iH(t)+\sum_{m=1}^{M}L_{m}^{\dagger}(t)L_{m}(t)/2}\rho\Bigr]-1,\label{eq:TDXi_Lindblad_def}
\end{align}
where $\mathbb{T}$ and $\overline{\mathbb{T}}$ are time-ordering and anti-time-ordering operators, respectively. Equations~\eqref{eq:Xi_Lindblad_def}
and \eqref{eq:TDXi_Lindblad_def} are the second main result of the
present Letter. Equation~\eqref{eq:main_result} with Eq.~\eqref{eq:TDXi_Lindblad_def}
is satisfied for the continuous measurement of jump events in any
Lindblad equation starting from any initial density operator.

Equations \eqref{eq:Jump_V0_def} and \eqref{eq:Jump_Vc_def} are
not the only Kraus representations compatible with the Lindblad equation
\eqref{eq:Lindblad_def}. A Kraus operator $Y_{m}$ compatible with
Eq.~\eqref{eq:Lindblad_def} (i.e., $\sum_{m}X_{m}\rho X_{m}^{\dagger}=\sum_{m}Y_{m}\rho Y_{m}^{\dagger}$)
can be obtained by $Y_{m^{\prime}}=\sum_{m}J_{m^{\prime}m}X_{m}$,
where $J_{m^{\prime}m}$ is an arbitrary unitary operator. This unitary
freedom of the Kraus operator corresponds to that in the measurement
basis of the environment. As mentioned above, $X_{m}$ is obtained
by the measurement basis $\ket{m}$ for each time interval, that is,
$X_{m}=\braket{m|U|0}$, while $Y_{m'}$ is derived via a different
measurement basis $\ket{\varphi_{m^{\prime}}}\equiv\sum_{m=0}^{M}(J^{\dagger})_{mm^{\prime}}\ket{m}$,
specifically, $Y_{m^{\prime}}=\braket{\varphi_{m^{\prime}}|U|0}$.
$\Xi$ in Eq.~\eqref{eq:Xi_Lindblad_def} depends on how we measure
the environment, that is, how we unravel the Lindblad equation. To
observe the consequences of different unravellings, for simplicity,
we consider a case having only one jump operator $L$. The Lindblad
equation is invariant under the following transformation: $H\to H-\frac{i}{2}(\zeta^{*}L-\zeta L^{\dagger})$
and $L\to L+\zeta\mathbb{I}_{S}$, where $\zeta\in\mathbb{C}$ is
an arbitrary parameter. A physical interpretation of this transformation
is presented in Refs.~\cite{Wiseman:1993:QJump,Kist:1999:SSE,Santos:2011:SSE}.
Under this transformation, $\Xi$ becomes (for time-independent $L$
and $H$) $\Xi=e^{|\zeta|^{2}T}\mathrm{Tr}_{S}\left[e^{T(iH+\frac{1}{2}L^{\dagger}L+\zeta^{*}L)}e^{T(-iH+\frac{1}{2}L^{\dagger}L+\zeta L^{\dagger})}\rho\right]-1$.
Therefore, for $|\zeta|\to\infty$, $\Xi$ scales as $\Xi\sim e^{|\zeta|^{2}T}$;
this indicates that we can make the lower bound of Eq.~\eqref{eq:main_result}
arbitrarily small by employing a continuous measurement with a large
$|\zeta|$. This result may appear contradictory to that obtained
in Ref.~\cite{Hasegawa:2020:QTURPRL}, which reported a unified lower
bound valid for any continuous measurements. Note that the continuous
measurements considered in Ref.~\cite{Hasegawa:2020:QTURPRL} require
a scaling condition, which is not satisfied for the measurements corresponding
to the transformation above.

\emph{Classical Markov processes.}---When we emulate classical Markov
processes with the Lindblad equation, $[H,\sum_{m=1}^{M}L_{m}^{\dagger}L_{m}]=0$
holds. In this case, from Eq.~\eqref{eq:Xi_Lindblad_def}, we obtain
\begin{equation}
\Xi_{\mathrm{CL}}=\mathrm{Tr}_{S}\left[e^{T\sum_{m=1}^{M}L_{m}^{\dagger}L_{m}}\rho\right]-1,\label{eq:Xi_classical_def}
\end{equation}
where the subscript ``CL'' is shorthand for ``classical.'' Therefore,
noncommutativeness $[H,\sum_{m=1}^{M}L_{m}^{\dagger}L_{m}]\ne0$ can
be a benefit of the quantum systems over their classical counterparts.
We evaluate the effect of noncommutativeness in the survival activity.
Assuming that $T$ is sufficiently small, a simple calculation yields
\cite{Supp:PhysRev} 
\begin{equation}
\Xi=\Xi_{\mathrm{CL}}+\frac{1}{2}T^{2}\chi+O(T^{3}),\label{eq:Xi_diff_def}
\end{equation}
where $\chi\equiv i\sum_{m=1}^{M}\mathrm{Tr}_{S}\left[[H,L_{m}^{\dagger}L_{m}]\rho\right]$
represents the expectation of the commutative relation. When $\chi>0$,
the system gains a precision enhancement due to its quantum nature. 

As a corollary of the continuous measurement, we can obtain a specific
expression of $\Xi_{\mathrm{CL}}$ for classical Markov processes.
We consider a classical Markov process with $N_{S}$ states $\{B_{1},B_{2},...,B_{N_{S}}\}$
and a transition rate $\gamma_{ji}(t)$ corresponding to a jump from
$B_{i}$ to $B_{j}$ at time $t$. Suppose that the initial probability
at state $B_{i}$ is given by $P_{i}$ ($\sum_{i=1}^{N_{S}}P_{i}=1$
and $P_{i}\ge0$). Then, Eq.~\eqref{eq:Xi_classical_def} is expressed
as 
\begin{equation}
\Xi_{\mathrm{CL}}=\sum_{i=1}^{N_{S}}\frac{P_{i}}{\mathcal{R}_{i}(T)}-1,\label{eq:Xi_classical}
\end{equation}
where $\mathcal{R}_{i}(T)\equiv e^{-\int_{0}^{T}dt\,\sum_{j\ne i}\gamma_{ji}(t)}$
is the survival probability in which there is no jump during $[0,T]$
starting from ${B_{i}}$. In Eq.~\eqref{eq:Xi_classical}, the first
term is the reciprocal expectation of the survival probability, which
is an experimentally measurable quantity. For the classical Markov
process, a classical representation of the counting observable $\mathcal{G}$
becomes $\mathcal{G}=\sum_{i,j,i\ne j}G_{ji}\mathcal{N}_{ji}$, where
$G_{ji}\in\mathbb{R}$ is a weight for the jump from $B_{i}$ to $B_{j}$
and $\mathcal{N}_{ji}$ is the number of jumps from $B_{i}$ to $B_{j}$
during $[0,T]$. Equation~\eqref{eq:main_result} with Eq.~\eqref{eq:Xi_classical}
is satisfied for arbitrary time-dependent Markov processes and initial
states. When the system activity is greater, $\mathcal{R}_{i}(T)$
decreases, resulting in a smaller lower bound. Indeed, for a short
time limit $T\to0$, $\Xi_{\mathrm{CL}}$ reduces to $\Xi_{\mathrm{CL}}\to\Upsilon$,
where $\Upsilon$ is the dynamical activity $\Upsilon\equiv\sum_{i,j,i\ne j}\int_{0}^{T}P_{i}(t)\gamma_{ji}(t)dt$.
Here, $P_{i}(t)$ is the probability of being $B_{i}$ at time $t$.
The dynamical activity quantifies the average number of jumps during
$[0,T]$. In classical Markov processes, the dynamical activity has
been reported to constitute the bound in the TUR \cite{Garrahan:2017:TUR,Terlizzi:2019:KUR,Vu:2019:UTURPRE}
and the QSL \cite{Shiraishi:2018:SpeedLimit}. For a steady-state
condition, it has been reported that the fluctuations in counting
observables are bounded from below by $1/\Upsilon$ \cite{Garrahan:2017:TUR}.
However, as demonstrated numerically in \cite{Supp:PhysRev}, in some
cases, $\mathrm{Var}[\mathcal{G}]/\braket{\mathcal{G}}^{2}\ge1/\Upsilon$
does not hold when the system is far from a steady state \cite{Supp:PhysRev}.

\emph{Quantum walk.}---We apply the main result of Eq.~\eqref{eq:main_result}
to a discrete-time one-dimensional quantum walk \cite{Ambainis:2001:QWalk,Venegas:2012:QWalkReview}.
The quantum walk is defined on the chirality space spanned by $\{\ket{\mathsf{R}},\ket{\mathsf{L}}\}$
and the position space spanned by $\{\ket{n}\}$, where $n$ is an
integer. Here, we identify the chirality and position spaces as the
principal and environmental systems, respectively (Fig.~\ref{fig:input_output}(c)).
One-step evolution of the quantum walk is performed via the unitary
operator $\mathcal{U}\equiv\mathcal{S}(\mathcal{K}\otimes\mathbb{I}_{E})$,
where $\mathbb{I}_{E}$ is an identity operator in $E$ and $\mathcal{K}$
and $\mathcal{S}$ are the coin and conditional shift operators, respectively.
For the coin operator, we employ the Hadamard gate defined by 
\begin{equation}
\mathcal{K}=\frac{1}{\sqrt{2}}\left(\ket{\mathsf{R}}\bra{\mathsf{R}}+\ket{\mathsf{R}}\bra{\mathsf{L}}+\ket{\mathsf{L}}\bra{\mathsf{R}}-\ket{\mathsf{L}}\bra{\mathsf{L}}\right).\label{eq:coinop_def}
\end{equation}
The conditional shift operator is given by 
\begin{equation}
\mathcal{S}=\sum_{n}\left[\ket{\mathsf{R}}\bra{\mathsf{R}}\otimes\ket{n+1}\bra{n}+\ket{\mathsf{L}}\bra{\mathsf{L}}\otimes\ket{n-1}\bra{n}\right],\label{eq:shiftop_def}
\end{equation}
which increases the position when the chirality is $\ket{\mathsf{R}}$
and decreases it when the chirality is $\ket{\mathsf{L}}$. The composite
system after $t$ steps is given by $\ket{\Psi(t)}=\mathcal{U}^{t}\ket{\Psi(0)}$,
where $\ket{\Psi(0)}$ is the initial state $\ket{\Psi(0)}=\ket{\mathsf{R}}\otimes\ket{0}$.
By using the combinatorics, the amplitudes at step $t$ can be computed
\cite{Meyer:1996:QCA,Ambainis:2001:QWalk,Brun:2003:QWMany}. At step
$t=T$, the measurement is performed on the position space, where
the measurement operator is defined by $\sum_{n}g(n)\ket{n}\bra{n}$.
Typically, $g(n)=n$ is employed, which corresponds to measuring the
position after $T$ steps. When $g(n)$ satisfies Eq.~\eqref{eq:F0_vanish},
that is, $g(n)$ is a counting observable, Eq.~\eqref{eq:main_result}
holds. Then, we obtain 
\begin{equation}
\Xi=\begin{cases}
2^{2u+1}\binom{u}{\frac{u}{2}}^{-2}-1 & u\in\mathrm{even},\\
2^{2u-1}\binom{u-1}{\frac{u-1}{2}}^{-2}-1 & u\in\mathrm{odd},
\end{cases}\label{eq:Xi_quantum_walk_def}
\end{equation}
where $u\equiv T/2$. Note that we only consider even $T$ because
the amplitudes vanish for odd $T$. Using Stirling's approximation,
$2^{2u+1}\binom{u}{\frac{u}{2}}^{-2}\sim\pi u$, indicating that the
survival activity linearly depends on the number of steps. This is
in contrast to the classical case where $\Xi$ exponentially depends
on time {[}see Eq.~\eqref{eq:Xi_classical}{]}. Although the environment
confers qualitatively different information in the continuous measurement
(Fig.~\ref{fig:input_output}(b)) and in the quantum walk (Fig.~\ref{fig:input_output}(c)),
our result can provide the lower bounds for both systems in a unified
way.

We also test the main result numerically for both classical and quantum
systems to verify the bound \cite{Supp:PhysRev}.

\emph{Derivation.}---We provide a brief derivation of Eq.~\eqref{eq:main_result}
(see \cite{Supp:PhysRev} for details). Our derivation is based on
the \emph{quantum} Cram\'er--Rao inequality \cite{Helstrom:1976:QuantumEst,Hotta:2004:QEstimation,Paris:2009:QFI,Liu:2019:QFisherReviewJPA},
which has been used to derive the QSL \cite{Jones:2010:QSL,Frowis:2012:FisherQSL,Taddei:2013:QSL}
and the TUR \cite{Hasegawa:2020:QTURPRL}. Suppose that the system
evolves according to Eq.~\eqref{eq:Psi_evolution}, in which $U$
and $V_{m}$ ($0\le m\le M$) are parametrized by a real parameter
$\theta$ as $U(\theta)$ and $V_{m}(\theta)$, respectively. The
final state of $S+E$ depends on $\theta$, which is expressed as
$\ket{\Psi_{\theta}(T)}$. For arbitrary measurement operator $\Theta_{E}$
in $E$, the quantum Cram\'er--Rao inequality holds \cite{Hotta:2004:QEstimation}:
\begin{equation}
\frac{\mathrm{Var}_{\theta}[\Theta_{E}]}{\left[\partial_{\theta}\braket{\Theta_{E}}_{\theta}\right]^{2}}\ge\frac{1}{\mathcal{F}_{E}(\theta)},\label{eq:QCRB_def}
\end{equation}
where $\mathcal{F}_{E}(\theta)$ is a quantum Fisher information \cite{Paris:2009:QFI,Liu:2019:QFisherReviewJPA},
$\braket{\Theta_{E}}_{\theta}\equiv\braket{\Psi_{\theta}(T)|\mathbb{I}_{S}\otimes\Theta_{E}|\Psi_{\theta}(T)}$,
and $\mathrm{Var}_{\theta}[\Theta_{E}]=\braket{\Theta_{E}^{2}}_{\theta}-\braket{\Theta_{E}}_{\theta}^{2}$.
From Ref.~\cite{Escher:2011:NoisyQFI}, $\mathcal{F}_{E}(\theta)$
is bounded from above by $\mathcal{F}_{E}(\theta)\le\mathcal{C}(\theta)$,
where $\mathcal{C}(\theta)\equiv4\left[\braket{\psi|H_{1}(\theta)|\psi}-\braket{\psi|H_{2}(\theta)|\psi}^{2}\right]$
with $H_{1}(\theta)\equiv\sum_{m=0}^{M}\left(\partial_{\theta}V_{m}^{\dagger}(\theta)\right)\left(\partial_{\theta}V_{m}(\theta)\right)$
and $H_{2}(\theta)\equiv i\sum_{m=0}^{M}\left(\partial_{\theta}V_{m}^{\dagger}(\theta)\right)V_{m}(\theta)$.

To derive the main result {[}Eq.~\eqref{eq:main_result}{]}, for
$1\le m\le M$, we consider the parametrization $V_{m}(\theta)\equiv e^{\theta/2}V_{m}$,
where $\theta=0$ recovers the original operator. Because a completeness
relation should be satisfied, $V_{0}(\theta)$ obeys $V_{0}^{\dagger}(\theta)V_{0}(\theta)=\mathbb{I}_{S}-\sum_{m=1}^{M}V_{m}^{\dagger}(\theta)V_{m}(\theta)=\mathbb{I}_{S}-e^{\theta}\sum_{m=1}^{M}V_{m}^{\dagger}V_{m}$.
For any $V_{0}(\theta)$ satisfying the completeness relation, there
exist a unitary operator $U_{V}$ such that $V_{0}(\theta)=U_{V}\sqrt{\mathbb{I}_{S}-e^{\theta}\sum_{m=1}^{M}V_{m}^{\dagger}V_{m}}$.
Substituting $V_{m}(\theta)$ into $\mathcal{C}(\theta)$ (as detailed
in \cite{Supp:PhysRev}), we find 
\begin{equation}
\mathcal{C}(\theta)=\braket{\psi|(V_{0}^{\dagger}V_{0})^{-1}|\psi}-1.\label{eq:FQ_Psi}
\end{equation}

We next evaluate $\braket{\mathcal{G}}_{\theta}$.
Because we have assumed that $g(0)=0$ {[}Eq.~\eqref{eq:F0_vanish}{]},
the complicated scaling dependence of $V_{0}(\theta)$ on $\theta$
can be ignored when computing $\braket{\mathcal{G}}_{\theta}$. Specifically,
we obtain $\braket{\mathcal{G}}_{\theta}=e^{\theta}\braket{\mathcal{G}}_{\theta=0}$.
We evaluate Eq.~\eqref{eq:QCRB_def} at $\theta=0$ with $\Theta_{E}=\mathcal{G}$
to obtain the main result {[}Eq.~\eqref{eq:main_result}{]}. Although
the derivation described here assumes that the initial state of $S$
is pure (i.e., $\rho=\ket{\psi}\bra{\psi}$), we can show that Eq.~\eqref{eq:main_result}
still holds for any initial mixed state $\rho$ in $S$ \cite{Supp:PhysRev}.

\emph{Conclusion.}---In this letter, we have derived a TUR for open
quantum systems. Because our relation holds for general open quantum
system, we expect the present study to serve as a basis for obtaining
the thermodynamic bound for several quantum systems, such as quantum
computation and communication.
\begin{acknowledgments}
This work was supported by the Ministry of Education, Culture, Sports,
Science and Technology (MEXT) KAKENHI Grant No.~JP19K12153. 
\end{acknowledgments}

\end{document}


\title{Supplementary Material for \\``Thermodynamic Uncertainty Relation for General Open Quantum Systems''}
\author{Yoshihiko Hasegawa}
\email{hasegawa@biom.t.u-tokyo.ac.jp}
\affiliation{Department of Information and Communication Engineering, Graduate
School of Information Science and Technology, The University of Tokyo,
Tokyo 113-8656, Japan}

\maketitle
This supplementary material provides more detail on the calculations introduced in the main text.  Equation and figure numbers are prefixed with S (e.g., Eq.~(S1) or Fig.~S1). Numbers without this prefix (e.g., Eq.~(1) or Fig.~1) refer to items in the main text.

\section{Derivation\label{sec:derivation}}

\subsection{Initially pure state case}
Our derivation is based on the \emph{quantum} Cram\'er--Rao inequality \cite{Helstrom:1976:QuantumEst,Hotta:2004:QEstimation,Paris:2009:QFI,Liu:2019:QFisherReviewJPA}, which has been used previously to derive the QSL \cite{Jones:2010:QSL,Frowis:2012:FisherQSL,Taddei:2013:QSL} and the TUR \cite{Hasegawa:2020:QTURPRL}. We first show the main result for initially pure states. Suppose that the system evolves according to Eq.~\PsiUevolution{}, where $U$ and $V_m$ ($0\le m \le M$) are parametrized by $\theta \in \mathbb{R}$ as $U(\theta)$ and $V_m(\theta)$, respectively. 
Then, the final state $\ket{\Psi(T)}$ depends on $\theta$, which is expressed by $\ket{\Psi_\theta(T)}$. 
Reference~\cite{Escher:2011:NoisyQFI} showed that the quantum Fisher information \cite{Paris:2009:QFI,Liu:2019:QFisherReviewJPA} of the principal system $\mathcal{F}_S(\theta)$ is bounded from above by
\begin{equation}
\mathcal{F}_S(\theta) = \max_{\mathcal{M}_S} F(\theta; \mathcal{M}_S\otimes \mathbb{I}_E) \le 
\max_{\mathcal{M}_{SE}} F(\theta; \mathcal{M}_{SE}) \equiv \mathcal{C}(\theta),
\label{eq:Fisher_upper_bound}
\end{equation}where $\mathcal{M}_S$ and $\mathcal{M}_{SE}$ are positive operator-valued measures (POVMs) in $S$ and $S+E$, respectively; $F(\theta; \mathcal{M})$ is a classical Fisher information obtained with $\mathcal{M}$; and $\mathbb{I}_E$ is the identity operator in $E$. $\mathcal{C}(\theta)$ is represented by \cite{Escher:2011:NoisyQFI}
\begin{equation}
\mathcal{C}(\theta) = 4\left[\braket{\psi|H_{1}(\theta)|\psi}-\braket{\psi|H_{2}(\theta)|\psi}^{2}\right],
\label{eq:CQ_def}
\end{equation}where $H_1(\theta)$ and $H_2(\theta)$ are
\begin{align}
H_{1}(\theta)&\equiv\sum_{m=0}^{M}\frac{dV_{m}^{\dagger}(\theta)}{d\theta}\frac{dV_{m}(\theta)}{d\theta},\label{eq:H1_def}\\
H_{2}(\theta)&\equiv i\sum_{m=0}^{M}\frac{dV_{m}^{\dagger}(\theta)}{d\theta}V_{m}(\theta).
\label{eq:H2_def}
\end{align}

Rather than considering the parameter inference in the principal system $S$, we consider the inference in the environment $E$. Let $\mathcal{F}_E(\theta)$ be the quantum Fisher information in $E$:
\begin{equation}
\mathcal{F}_E(\theta) \equiv \max_{\mathcal{M}_E} F(\theta; \mathbb{I}_S \otimes \mathcal{M}_E),
\label{eq:FE_def}
\end{equation}where $\mathcal{M}_E$ is a POVM in $E$. Let $\Theta_E$ be an arbitrary observable in $E$. Then, according to the quantum Cram\'er--Rao inequality, we have
\begin{equation}
\frac{\mathrm{Var}_{\theta}[\Theta_E]}{\left[\partial_{\theta}\braket{\Theta_E}_{\theta}\right]^{2}}\ge\frac{1}{\mathcal{F}_{E}(\theta)} \ge \frac{1}{\mathcal{C}(\theta)},
\label{eq:QCRBE2}
\end{equation}where $\braket{\Theta_E}_\theta \equiv \braket{\Psi_\theta(T) | \mathbb{I}_S\otimes \Theta_E | \Psi_\theta(T)}$. 
The second inequality originates from $\mathcal{F}_E(\theta) = \max_{\mathcal{M}_E} F(\theta; \mathbb{I}_S \otimes \mathcal{M}_E) \le \max_{\mathcal{M}_{SE}} F(\theta; \mathcal{M}_{SE})=\mathcal{C}(\theta)$.

Because Eq.~\eqref{eq:QCRBE2} holds for arbitrary $\Theta_E$ in $E$, substituting $\Theta_E = \mathcal{G}$ yields
\begin{equation}
    \frac{\mathrm{Var}_{\theta}[\mathcal{G}]}{\left[\partial_{\theta}\braket{\mathcal{G}}_{\theta}\right]^{2}}\ge \frac{1}{\mathcal{C}(\theta)}.\label{eq:QCRB2}
\end{equation}

To derive the main result [Eq.~\mainUresult{}] from Eq.~\eqref{eq:QCRB2}, we consider the following parametrization for $V_{m\ge 1}(\theta)$:
\begin{equation}
V_{m}(\theta)\equiv e^{\theta/2}V_{m}\;\;\;(1\le m\le M),\label{eq:Vm_theta_def}
\end{equation}where $\theta=0$ recovers the original operator. 
Note that we cannot scale $V_0(\theta)$ as $e^{\theta/2}V_0$ due to the completeness relation $\sum_{m=0}^M V_m^\dagger(\theta)V_m(\theta) = \mathbb{I}_S$. 
To satisfy the completeness relation, $V_{0}(\theta)$ obeys $V_{0}^{\dagger}(\theta)V_{0}(\theta)=\mathbb{I}_S-\sum_{m=1}^{M}V_{m}^{\dagger}(\theta)V_{m}(\theta)=\mathbb{I}_S-e^{\theta}\sum_{m=1}^{M}V_{m}^{\dagger}V_{m}$. 
This yields
\begin{equation}
V_{0}^{\dagger}(\theta)V_{0}(\theta)=e^{\theta}V_{0}^{\dagger}(0)V_{0}(0)-[e^{\theta}-1]\mathbb{I}_{S},
\label{eq:V0_contraint}
\end{equation}where we find that the last part of  the right-hand side is not positive definite for $\theta > 0$. Therefore the positive definiteness of $V_0^\dagger(\theta)V_0(\theta)$ demands that $V_0^\dagger (0) V_0(0)$ be positive definite, indicating that $V_0(0)=V_0$ should be a full-rank matrix.

For any $V_{0}(\theta)$ satisfying the completeness relation, there exists a unitary operator $U_V$ such that 
\begin{equation}
V_{0}(\theta)=U_V\sqrt{\mathbb{I}_S-e^{\theta}\sum_{m=1}^{M}V_{m}^{\dagger}V_{m}}.\label{eq:V0_theta_def}
\end{equation}Because $\sum_{m=1}^{M}V_m^\dagger V_m$ is Hermitian, we can consider the spectral decomposition of $\sum_{m=1}^{M}V_m^\dagger V_m$:
\begin{equation}
\sum_{m=1}^{M}V_{m}^{\dagger}V_{m}=\sum_{i}\zeta_{i}\Pi_{i},
\label{eq:spectral_decomposition}
\end{equation}where $\{\zeta_i\}$ is a set of different eigenvalues of $\sum_{m=1}^{M}V_m^\dagger V_m$ and $\Pi_i$ is a projector corresponding to $\zeta_i$. Given a univariate complex function $f(x)$, an operator function $f(A)$, where $A$ is an Hermitian operator, is defined by $f(A)\equiv \sum_i f(\zeta_i) \Pi_i$ \cite{Nielsen:2011:QuantumInfoBook}. Then, Eq.~\eqref{eq:V0_theta_def} is given by
\begin{equation}
V_{0}(\theta)=U_V\sum_{i}\sqrt{1-e^{\theta}\zeta_{i}}\Pi_{i}.\label{eq:V0_decomp_def}
\end{equation}

First, we calculate $H_1(\theta)$ defined by Eq.~\eqref{eq:H1_def}. We substitute Eqs.~\eqref{eq:V0_decomp_def} and \eqref{eq:Vm_theta_def} into Eq.~\eqref{eq:H1_def} to obtain
\begin{align}
H_{1}(\theta=0)&=\frac{1}{4}\sum_{m=1}^{M}V_{m}^{\dagger}V_{m}+\left(\frac{d}{d\theta}V_{0}^{\dagger}(\theta)\right)\left(\frac{d}{d\theta}V_{0}(\theta)\right)_{\theta=0}\nonumber\\&=\frac{1}{4}\sum_{i}\zeta_{i}\Pi_{i}+\frac{1}{4}\sum_{i}\frac{\zeta_{i}^{2}}{1-\zeta_{i}}\Pi_{i}\nonumber\\&=\frac{1}{4}\sum_{i}\frac{\zeta_{i}}{1-\zeta_{i}}\Pi_{i}\nonumber\\&=\frac{1}{4}\left(\sum_{m=1}^{M}V_{m}^{\dagger}V_{m}\right)\left(\mathbb{I}_S-\sum_{m=1}^{M}V_{m}^{\dagger}V_{m}\right)^{-1}\nonumber\\&=\frac{1}{4}\left[(V_{0}^{\dagger}V_{0})^{-1}-1\right].
\label{eq:H1_expr}
\end{align}
Similarly, $H_2(\theta)$ defined by Eq.~\eqref{eq:H2_def} is obtained as follows:
\begin{align}
H_{2}(\theta=0)&=i\left[\sum_{m=1}^{M}\frac{dV_{m}(\theta)^{\dagger}}{d\theta}V_{m}(\theta)+\frac{dV_{0}(\theta)^{\dagger}}{d\theta}V_{0}(\theta)\right]_{\theta=0}\nonumber\\&=i\left[\frac{1}{2}\sum_{m=1}^{M}V_{m}^{\dagger}V_{m}-\sum_{i,j}\frac{\zeta_{i}\sqrt{1-\zeta_{j}}}{2\sqrt{1-\zeta_{i}}}\Pi_{i}\Pi_{j}\right]\nonumber\\&=i\left[\frac{1}{2}\sum_{m=1}^{M}V_{m}^{\dagger}V_{m}-\frac{1}{2}\sum_{i}\zeta_{i}\Pi_{i}\right]\nonumber\\&=0.
\label{eq:H2_expr}
\end{align}Equations~\eqref{eq:H1_expr} and \eqref{eq:H2_expr} are substituted into Eq.~\eqref{eq:CQ_def} to give $\mathcal{C}(\theta=0)$ as 
\begin{equation}
\mathcal{C}(\theta=0)=\braket{\psi|(V_{0}^{\dagger}V_{0})^{-1}|\psi}-1.
\label{eq:FQ_Psi}
\end{equation}

We next evaluate $\braket{\mathcal{G}}_{\theta}$ in Eq.~\eqref{eq:QCRB2}. As we have assumed that $g(0)=0$ [Eq.~\FOUvanish{}], the complicated scaling dependence of $V_{0}(\theta)$ on $\theta$ [i.e., Eq.~\eqref{eq:V0_theta_def}] can be ignored when computing $\braket{\mathcal{G}}_\theta$. Specifically, we obtain
\begin{align}
\braket{\mathcal{G}}_{\theta}&=\braket{\Psi_{\theta}(T)|\mathbb{I}_{S}\otimes\mathcal{G}|\Psi_{\theta}(T)}\nonumber\\&=\left(\sum_{m}\bra{\psi_{S}}V_{m}^{\dagger}(\theta)\otimes\bra{\phi_{m}}\right)\left(\mathbb{I}_{S}\otimes\sum_{m'}g(m')\ket{\phi_{m'}}\bra{\phi_{m'}}\right)\left(\sum_{m''}V_{m''}(\theta)\ket{\psi_{S}}\otimes\ket{\phi_{m''}}\right)\nonumber\\&=\sum_{m=0}^{M}g(m)\bra{\psi_{S}}V_{m}^{\dagger}(\theta)V_{m}(\theta)\ket{\psi_{S}}\nonumber\\&=\sum_{m=1}^{M}g(m)\bra{\psi_{S}}V_{m}^{\dagger}(\theta)V_{m}(\theta)\ket{\psi_{S}}\nonumber\\&=e^{\theta}\braket{\mathcal{G}}_{\theta=0}.
 \label{eq:g_scaling}
\end{align}We evaluate Eq.~\eqref{eq:QCRB2} at $\theta=0$ with Eqs.~\eqref{eq:FQ_Psi} and \eqref{eq:g_scaling} to obtain the main result as Eq.~\mainUresult{}. 

\subsection{Initially mixed-state case}
We have derived the main result [Eq.~\mainUresult{}] for initially pure states. Here, we show that the main result still holds for initially mixed states through a purification. 

First, we introduce an ancilla $S^\prime$ that purifies a mixed state in $S$. Let $\ket{\tilde{\psi}}$ be a pure state in $S+S^\prime$, which is a purification of $\rho$:
\begin{equation}
\rho=\mathrm{Tr}_{S^{\prime}}\left[\ket{\tilde{\psi}}\bra{\tilde{\psi}}\right].
\label{eq:purification}
\end{equation}We can describe the time evolution of $\ket{\tilde{\psi}}\bra{\tilde{\psi}}$ as follows:
\begin{equation}
\tilde{\rho}(T) = \sum_{m=0}^{M}\tilde{V}_{m}\ket{\tilde{\psi}}\bra{\tilde{\psi}}\tilde{V}_{m}^{\dagger},
\label{eq:varrho_evo}
\end{equation}where $\tilde{V}_m$ is an operator in $S+S^\prime$ and $\tilde{\rho}(T)$ is a density operator in $S+S^\prime$ at time $t=T$. The pure state $\ket{\tilde{\psi}}$ in $S+S^\prime$ is used to obtain the pure state of $S+S^\prime +E$ after the interaction: 
\begin{equation}
\ket{\tilde{\Psi}(T)} = \sum_{m=0}^{M} \tilde{V}_m \ket{\tilde{\psi}} \otimes \ket{\phi_m}.
\label{eq:PhiT_def}
\end{equation}$\tilde{V}_m$ should be defined such that it yields a consistent evolution for $\rho(T)$ in $S$, that is, 
\begin{align}
\rho(T)&=\mathrm{Tr}_{S^{\prime}}\left[\sum_{m=0}^{M}\tilde{V}_{m}\ket{\tilde{\psi}}\bra{\tilde{\psi}}\tilde{V}_{m}^{\dagger}\right]\\&=\sum_{m=0}^{M}V_{m}\rho V_{m}^{\dagger}.
\label{eq:rho_evo_purify}
\end{align}Equation~\eqref{eq:PhiT_def} is parametrized as
\begin{equation}
\ket{\tilde{\Psi}_\theta(T)} = \sum_{m=0}^{M} \tilde{V}_m(\theta) \ket{\tilde{\psi}} \otimes \ket{\phi_m}.
\label{eq:PhiT_theta_def}
\end{equation}Reference~\cite{Escher:2011:NoisyQFI} showed that the upper bound on the quantum Fisher information after purification is
\begin{align}
\mathcal{F}_E(\theta) &\le \tilde{\mathcal{C}}(\theta),\label{eq:FC_mixed}\\
\tilde{\mathcal{C}}(\theta) & \equiv 4\left[\mathrm{Tr}_{S}[H_{1}(\theta)\rho]-\mathrm{Tr}_{S}[H_{2}(\theta)\rho]^{2}\right],\label{eq:FQ_purify}
\end{align}where $H_1(\theta)$ and $H_2(\theta)$ are defined in Eqs.~\eqref{eq:H1_def} and \eqref{eq:H2_def}, respectively

We next evaluate $\braket{\mathcal{G}}_{\theta}=\braket{\tilde{\Psi}_{\theta}(T)|\mathbb{I}_{S}\otimes\mathbb{I}_{S^{\prime}}\otimes\mathcal{G}|\tilde{\Psi}_{\theta}(T)}$ in Eq.~\eqref{eq:QCRB2}. Specifically, we obtain
\begin{align}
\braket{\mathcal{G}}_{\theta}&=\mathrm{Tr}_{SS^{\prime}E}\left[\left(\mathbb{I}_{S}\otimes\mathbb{I}_{S^{\prime}}\otimes\mathcal{G}\right)\ket{\tilde{\Psi}_{\theta}(T)}\bra{\tilde{\Psi}_{\theta}(T)}\right]\nonumber\\&=\mathrm{Tr}_{SS^{\prime}E}\Biggl[\left(\mathbb{I}_{S}\otimes\mathbb{I}_{S^{\prime}}\otimes\mathcal{G}\right)\left(\sum_{m=0}^{M}\sum_{m^{\prime}=0}^{M}\tilde{V}_{m}(\theta)\ket{\tilde{\psi}}\otimes\ket{\phi_{m}}\bra{\tilde{\psi}}\tilde{V}_{m^{\prime}}(\theta)\otimes\bra{\phi_{m^{\prime}}}\right)\Biggr]\nonumber\\&=\sum_{m=0}^{M}\sum_{m^{\prime}=0}^{M}\bra{\phi_{m^{\prime}}}\mathcal{G}\ket{\phi_{m}}\mathrm{Tr}_{SS^{\prime}}\left[\tilde{V}_{m}(\theta)\ket{\tilde{\psi}}\bra{\tilde{\psi}}\tilde{V}_{m^{\prime}}(\theta)\right]\nonumber\\&=\sum_{m=0}^{M}g(m)\mathrm{Tr}_{S}\left[V_{m}(\theta)\rho V_{m}^{\dagger}(\theta)\right]\nonumber\\&=e^{\theta}\braket{\mathcal{G}}_{\theta=0}.
 \label{eq:g_purify_scaling}
\end{align}Evaluating Eq.~\eqref{eq:QCRB2} at $\theta=0$, we obtain the main result [Eq.~\mainUresult{}] for initially mixed states. 

\subsection{Multidimensional generalization}

We can obtain a multivariate generalization of Eq.~\mainUresult{}. Let $\Theta_E$ and $\Phi_E$ be arbitrary commutable observables in $E$, with $[\Theta_E,\Phi_E] = 0$. Then, the multidimensional Cram\'er--Rao inequality holds:
\begin{equation}
\left[\begin{array}{cc}
\mathrm{Var}_{\theta}[\Theta_{E}] & \mathrm{Cov}_{\theta}[\Theta_{E},\Phi_{E}]\\
\mathrm{Cov}_{\theta}[\Theta_{E},\Phi_{E}] & \mathrm{Var}_{\theta}[\Phi_{E}]
\end{array}\right]-\frac{1}{\mathcal{F}_{E}(\theta)}\left[\begin{array}{c}
\partial_{\theta}\left\langle \Theta_{E}\right\rangle _{\theta}\\
\partial_{\theta}\left\langle \Phi_{E}\right\rangle _{\theta}
\end{array}\right]\left[\begin{array}{cc}
\partial_{\theta}\left\langle \Theta_{E}\right\rangle _{\theta} & \partial_{\theta}\left\langle \Phi_{E}\right\rangle _{\theta}\end{array}\right]\ge0,
\label{eq:multi_CRB}
\end{equation}where $\mathrm{Cov}_{\theta}[\Theta_{E},\Phi_{E}]\equiv\braket{\Theta_{E}\Phi_{E}}_{\theta}-\braket{\Theta_{E}}_{\theta}\braket{\Phi_{E}}_{\theta}$ is the covariance, and $A\ge 0$ denotes that $A$ is a positive semi-definite matrix.

As in the single-observable case, we assume that $S+E$ undergoes the unitary operator $U$, and measurement is applied to $E$ at the final state. Here, we consider two commutable operators $\mathcal{G}_1$ and $\mathcal{G}_2$, which are counting observables, for the measurement. From Eq.~\eqref{eq:multi_CRB} with Eqs.~\eqref{eq:FQ_Psi} and \eqref{eq:g_scaling}, the following relation holds:
\begin{equation}
\left[\begin{array}{cc}
\mathrm{Var}[\mathcal{G}_{1}] & \mathrm{Cov}[\mathcal{G}_{1},\mathcal{G}_{2}]\\
\mathrm{Cov}[\mathcal{G}_{1},\mathcal{G}_{2}] & \mathrm{Var}[\mathcal{G}_{2}]
\end{array}\right]-\frac{1}{\Xi}\left[\begin{array}{cc}
\braket{\mathcal{G}_{1}}^{2} & \braket{\mathcal{G}_{1}}\braket{\mathcal{G}_{2}}\\
\braket{\mathcal{G}_{1}}\braket{\mathcal{G}_{2}} & \braket{\mathcal{G}_{2}}^{2}
\end{array}\right]\ge0,
\label{eq:multi_TUR}
\end{equation}which is a multidimensional generalization of Eq.~\mainUresult{}. Generalization to more than two commutable observables is straightforward. 

\section{Calculations of survival activity\label{sec:calculation_of_xi}}
\subsection{Continuous measurement}

The survival activity is defined by Eq.~\XiUdef{}. Because $V_{0}$ is the action associated with no jump events within $[0,T]$, it is given by
\begin{equation}
V_{0}=\lim_{N\to\infty}X_{0}^{N}=\lim_{N\to\infty}\left(\mathbb{I}_{S}-i\Delta tH-\frac{1}{2}\Delta t\sum_{m=1}^{M}L_{m}^{\dagger}L_{m}\right)^{N}.
\label{eq:TI_V0_contmeas_def}
\end{equation}We use $\mathbb{I}_{S}-i\Delta tH-\frac{1}{2}\Delta t\sum_{m=1}^{M}L_{m}^{\dagger}L_{m}=\exp\left(-i\Delta tH-\frac{1}{2}\Delta t\sum_{m=1}^{M}L_{m}^{\dagger}L_{m}\right)+O((\Delta t)^{2})$ to calculate Eq.~\eqref{eq:TI_V0_contmeas_def} as
\begin{equation}
V_{0}=e^{-T\left(iH+\frac{1}{2}\sum_{m=1}^M L_{m}^{\dagger}L_{m}\right)}.
\label{eq:TI_V0_contmeas2}
\end{equation}Therefore, the survival activity $\Xi$ is
\begin{align}
\Xi&=\mathrm{Tr}_S\left[\left(e^{T\left(iH-\frac{1}{2}\sum_{m=1}^M L_{m}^{\dagger}L_{m}\right)}e^{T\left(-iH-\frac{1}{2}\sum_{m=1}^M L_{m}^{\dagger}L_{m}\right)}\right)^{-1}\rho\right]-1\nonumber\\&=\mathrm{Tr}_S\left[e^{T\left(iH+\frac{1}{2}\sum_{m=1}^M L_{m}^{\dagger}L_{m}\right)}e^{T\left(-iH+\frac{1}{2}\sum_{m=1}^M L_{m}^{\dagger}L_{m}\right)}\rho\right]-1.
\label{eq:TI_Xi_Lindblad_def}
\end{align}

When $H$ and $L_m$ depend on time, we can proceed in the same manner. In this case, $X_0(t)$ is expressed as
\begin{equation}
X_0(t) = \mathbb{I}_S - i \Delta t H(t) - \frac{1}{2}\Delta t \sum_{m=1}^M L_m^\dagger(t)L_m(t).
\label{eq:TD_X0_def}
\end{equation}For $\Delta t \to 0$, we have
\begin{equation}
\mathbb{I}_{S}-i\Delta tH(t)-\frac{1}{2}\Delta t\sum_{m=1}^{M}L_{m}^{\dagger}(t)L_{m}(t)=\exp\left[-i\Delta tH(t)-\frac{1}{2}\Delta t\sum_{m=1}^{M}L_{m}^{\dagger}(t)L_{m}(t)\right] + O((\Delta t)^2).
\label{eq:approx_with_exp}
\end{equation}Then, $V_{0}$ is
\begin{align}
V_{0}&=\lim_{N\to\infty}\left(X_{0}(t_{N-1})\cdots X_{0}(t_{1})X_{0}(t_{0})\right)\nonumber\\&=\lim_{N\to\infty}\left(e^{-i\Delta tH(t_{N-1})-\Delta t\sum_{m=1}^{M}L_{m}^{\dagger}(t_{N-1})L_{m}(t_{N-1})/2}\times\cdots\times e^{-i\Delta tH(t_{0})-\Delta t\sum_{m=1}^{M}L_{m}^{\dagger}(t_{0})L_{m}(t_{0})/2}\right).\label{eq:V0_contmeas_def}
\end{align}Let $A(t)$ be an arbitrary time-dependent matrix. The time-ordered and anti-time-ordered exponential operations are given by
\begin{align}
\mathbb{T}e^{\int_{0}^{T}dt\,A(t)}&=\lim_{N\to\infty}\left(e^{A(t_{N-1})\Delta t}\cdots e^{A(t_{1})\Delta t}e^{A(t_{0})\Delta t}\right),\label{eq:time_ordered_exp}\\
\overline{\mathbb{T}}e^{\int_{0}^{T}dt\,A(t)}&=\lim_{N\to\infty}\left(e^{A(t_{1})\Delta t}\cdots e^{A(t_{N-1})\Delta t}e^{A(t_{N})\Delta t}\right),\label{eq:rev_time_ordered_exp}
\end{align}where $\mathbb{T}$ and $\overline{\mathbb{T}}$ are time-ordering and anti-time-ordering operators, respectively. By using Eq.~\eqref{eq:time_ordered_exp}, Eq.~\eqref{eq:V0_contmeas_def} is
\begin{equation}
V_{0}=\mathbb{T}e^{\int_{0}^{T}dt\left(-iH(t)-\sum_{m=1}^M L_{m}^{\dagger}(t)L_{m}(t)/2\right)}.
\label{eq:V0_contmeas2}
\end{equation}The survival activity $\Xi$ is given by 
\begin{align}
\Xi&=\mathrm{Tr}_{S}\left[\overline{\mathbb{T}}e^{\int_{0}^{T}dt\,iH(t)+\sum_{m=1}^M L_{m}^{\dagger}(t)L_{m}(t)/2}\mathbb{T}e^{\int_{0}^{T}dt\,-iH(t)+\sum_{m=1}^M L_{m}^{\dagger}(t)L_{m}(t)/2}\rho\right]-1.
\label{eq:Xi_Lindblad_def}
\end{align}

A commutative relation $[H,\sum_{m=1}^M L_m^\dagger L_m]=0$ is assumed to calculate Eq.~\eqref{eq:TI_Xi_Lindblad_def} in
\begin{equation}
\Xi_{\mathrm{CL}}=\mathrm{Tr}_{S}\left[e^{T\sum_{m=1}^M L_{m}^{\dagger}L_{m}}\rho\right]-1.
\label{eq:Xi_classical_def_Supp}
\end{equation}Assume that $T$ is sufficiently small. Then, for any matrices $A$ and $B$, using $e^{TA} = \mathbb{I} + TA + \frac{1}{2}T^2 A^2 + O(T^3)$, we obtain
\begin{equation}
e^{TA}e^{TB} = e^{T(A+B)}+\frac{1}{2}T^{2}[A,B]+O(T^{3}).
\label{eq:expAB}
\end{equation}We substitute $A=i H + \frac{1}{2}\sum_{m=1}^M L_m^\dagger  L_m$ and $B=-i H + \frac{1}{2}\sum_{m=1}^M L_m^\dagger  L_m$ into Eq.~\eqref{eq:expAB} to obtain
\begin{equation}
e^{T\left(iH+\frac{1}{2}\sum_{m=1}^M L_{m}^{\dagger}L_{m}\right)}e^{T\left(-iH+\frac{1}{2}\sum_{m=1}^M L_{m}^{\dagger}L_{m}\right)} = e^{T\sum_{m=1}^M L_{m}^{\dagger}L_{m}}+\frac{i}{2}T^{2}\left [H,\sum_{m=1}^M L_{m}^{\dagger}L_{m}\right ]+O(T^{3}),
\label{eq:XiLE_XiCL}
\end{equation}which gives us Eq.~\XiUdiffUdef{} in the main text.

\subsection{Quantum walk}

As explained in the main text, the state of the composite system (system and environment) after $t$ steps is
\begin{equation}
\ket{\Psi(t)} = \mathcal{U}^t \ket{\Psi(0)},
\label{eq:QW_evolution}
\end{equation}where $\mathcal{U} \equiv \mathcal{S}(\mathcal{K} \otimes \mathbb{I}_E)$. Here, $\mathcal{K}$ and $\mathcal{S}$ are the coin and conditional shift operators, as defined in Eqs.~\coinopUdef{} and \shiftopUdef{}, respectively. By using the combinatorics, the amplitudes at step $t$ are given by \cite{Meyer:1996:QCA,Ambainis:2001:QWalk,Brun:2003:QWMany} 
\begin{align}
\braket{\mathsf{L},n|\mathcal{U}^{t}|\mathsf{R},0}&=\frac{1}{\sqrt{2^{t}}}\sum_{C=1}(-1)^{N_{\mathsf{L}}-C}\binom{N_{\mathsf{L}}-1}{C-1}\binom{N_{\mathsf{R}}}{C-1},\label{eq:LUR}\\\braket{\mathsf{R},n|\mathcal{U}^{t}|\mathsf{R},0}&=\frac{1}{\sqrt{2^{t}}}\sum_{C=1}(-1)^{N_{\mathsf{L}}-C}\binom{N_{\mathsf{L}}-1}{C-1}\binom{N_{\mathsf{R}}}{C},\label{eq:RUR}\\\braket{\mathsf{L},n|\mathcal{U}^{t}|\mathsf{L},0}&=\frac{1}{\sqrt{2^{t}}}\sum_{C=1}(-1)^{N_{\mathsf{L}}-C}\binom{N_{\mathsf{L}}-1}{C-1}\binom{N_{\mathsf{R}}}{C-1}\frac{N_{\mathsf{R}}-2C+2}{N_{\mathsf{R}}},\label{eq:LUL}\\\braket{\mathsf{R},n|\mathcal{U}^{t}|\mathsf{L},0}&=\frac{1}{\sqrt{2^{t}}}\sum_{C=1}(-1)^{N_{\mathsf{L}}-C}\binom{N_{\mathsf{L}}-1}{C-1}\binom{N_{\mathsf{R}}}{C}\frac{N_{\mathsf{R}}-2C}{N_{\mathsf{R}}},\label{eq:RUL}
\end{align}where $\binom{N_{1}}{N_{2}}$ is a binomial coefficient, $N_\mathsf{L}\equiv (t-n)/2$, and $N_\mathsf{R}\equiv (t+n)/2$. The upper bound of the summation in Eqs.~\eqref{eq:LUR}--\eqref{eq:RUL} is $N_\mathsf{L}$ for $n\ge0$ and $N_\mathsf{R}+1$ otherwise. Moreover, the boundary cases ($N_\mathsf{L}=0$ and $N_\mathsf{R}=0$) should be calculated separately. Because the boundary cases are irrelevant in our calculation, we do not show them here (see Ref.~\cite{Brun:2003:QWMany} for details). In the quantum walk case, $V_0=\braket{0|\mathcal{U}^T|0}$ is given by
\begin{equation}
V_{0}=\left[\begin{array}{cc}
\braket{\mathsf{R},0|\mathcal{U}^{T}|\mathsf{R},0} & \braket{\mathsf{R},0|\mathcal{U}^{T}|\mathsf{L},0}\\
\braket{\mathsf{L},0|\mathcal{U}^{T}|\mathsf{R},0} & \braket{\mathsf{L},0|\mathcal{U}^{T}|\mathsf{L},0}
\end{array}\right].
\label{eq:V0_QW_def}
\end{equation}After $n=0$ is substituted into Eqs.~\eqref{eq:LUR}--\eqref{eq:RUL}, $V_0$ in the basis of $\ket{\mathsf{R}}$ and $\ket{\mathsf{L}}$ is
\begin{equation}
V_{0}=\begin{cases}
{\displaystyle \frac{(-1)^{\frac{u}{2}}}{2^{u+1}}\binom{u}{\frac{u}{2}}}\left[\begin{array}{cc}
1 & 1\\
-1 & 1
\end{array}\right] & u\in\mathrm{even},\\
{\displaystyle \frac{(-1)^{\frac{u-1}{2}}}{2^{u}}\binom{u-1}{\frac{u-1}{2}}}\left[\begin{array}{cc}
1 & -1\\
1 & 1
\end{array}\right] & u\in\mathrm{odd},
\end{cases}
\label{eq:V0_QW2}
\end{equation}where $u\equiv T/ 2$. Note that we only consider even $T$ because for odd $T$, the amplitudes in Eq.~\eqref{eq:V0_QW_def} vanish, and accordingly, $V_0=0$. By using Eq.~\eqref{eq:V0_QW2}, we obtain
\begin{equation}
\Xi=\begin{cases}
{\displaystyle 2^{2u+1}\binom{u}{\frac{u}{2}}^{-2}-1} & u\in\mathrm{even},\\
{\displaystyle 2^{2u-1}\binom{u-1}{\frac{u-1}{2}}^{-2}-1} & u\in\mathrm{odd},
\end{cases}
\label{eq:Xi_quantum_walk_def}
\end{equation}which is Eq.~\XiUquantumUwalkUdef{} in the main text.

\section{Numerical verification}

\begin{figure*}
\includegraphics[width=18cm]{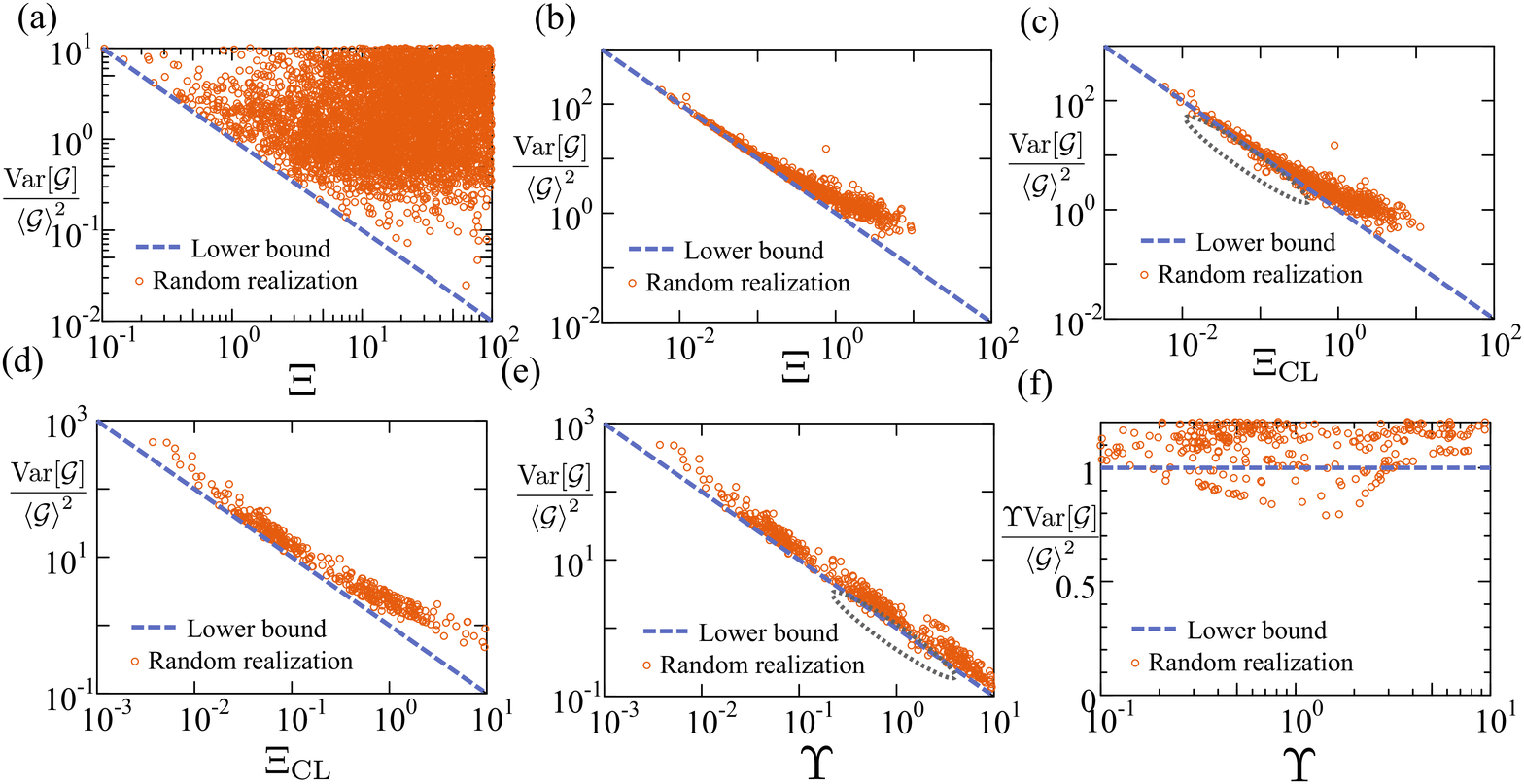}
\caption{Random realizations and their lower bound for different models. 
(a) Joint unitary evolution of system and environment. 
Circles denote $\mathrm{Var}[\mathcal{G}]/\braket{\mathcal{G}}^2$ as a function of $\Xi$ for random realizations, and the dashed line denotes $1/\Xi$.
The observable $\mathcal{G}$ is a random Hermitian matrix satisfying Eq.~\FOUvanish{}. 
The dimensionalities are randomly selected from $d_S\in\{2,3,...,10\}$ and $d_E\in\{2,3,...10\}$. 
(b) and (c) Photon counting in a two-level atom.
Circles denote $\mathrm{Var}[\mathcal{G}]/\braket{\mathcal{G}}^2$ as a function of (b) $\Xi$ and (c) $\Xi_\mathrm{CL}$ for random realizations, and the dashed lines are (b) $1/\Xi$ and (c) $1/\Xi_\mathrm{CL}$, where $\Xi$ and $\Xi_\mathrm{CL}$ are defined by Eqs.~\XiULindbladUdef{} and \XiUclassicalUdef{}, respectively.
In (c), circles inside the dotted ellipse are below the dashed line. 
Parameters are randomly selected from $\Delta \in [0.1,3.0]$, $\Omega \in [0.1,3.0]$, $\kappa \in [0.1,3.0]$, and $T\in[0.1, 1.0]$. 
(d)--(f) Classical time-dependent Markov process. 
Circles denote $\mathrm{Var}[\mathcal{G}]/\braket{\mathcal{G}}^2$ as a function of (d) $\Xi_\mathrm{CL}$ and (e) $\Upsilon$ for random realizations, and the dashed lines are (d) $1/\Xi_\mathrm{CL}$ and (e) $1/\Upsilon$, where $\Xi_\mathrm{CL}$ is defined by Eq.~\XiUclassical{}.
In (e), circles inside the dotted ellipse are below the dashed line. 
Circles in (f) denote $\Upsilon \mathrm{Var}[\mathcal{G}]/\braket{\mathcal{G}}^2$ as a function of $\Upsilon$ for random realizations, and the dashed line is $1$. 
Parameters are randomly selected from $T\in\{0.1,1.0,10.0\}$, $N_S\in \{3,4,...,6\}$, $\nu_{ji}\in [0,1]$, $A\in [0,1]$, $\omega\in[0.001,0.1]$, and $\vartheta\in[0,2\pi]$.  
\label{fig:bound_plot}}
\end{figure*}

\subsection{Joint unitary evolution of system and environment}

We verify the main result [Eq.~\mainUresult{}] with computer simulations. We numerically implement the joint unitary evolution depicted by Eq.~\PsiUevolution{} and the measurement on the environmental system after the evolution. Let $d_S$ and $d_E$ be the dimensionalities of the principal and the environmental systems, respectively.
We randomly generate the initial state $\ket{\psi}$, the unitary matrix $U$, and the observable $\mathcal{G}$, where $\mathcal{G}$ satisfies the condition of Eq.~\FOUvanish{} (the parameter ranges are shown in the caption of Fig.~\ref{fig:bound_plot}(a)). 
In Fig.~\ref{fig:bound_plot}(a), circles show $\mathrm{Var}[\mathcal{G}]/\braket{\mathcal{G}}^2$ as a function of $\Xi$ for many random realizations, where the dashed line denotes the lower bound of Eq.~\mainUresult{}. As seen in Fig.~\ref{fig:bound_plot}(a), all realizations are above the lower bound, which numerically verifies Eq.~\mainUresult{}.

\subsection{Quantum continuous measurement}
We next verify the main result obtained for the continuous measurement in the Lindblad equation. Specifically, we consider photon counting in a two-level atom driven by a classical laser field. Let $\ket{\epsilon_g}$ and $\ket{\epsilon_e}$ be the ground and excited states, respectively. The Hamiltonian and the jump operator are expressed as
\begin{align}
H&=\Delta\ket{\epsilon_{e}}\bra{\epsilon_{e}}+\frac{\Omega}{2}(\ket{\epsilon_{e}}\bra{\epsilon_{g}}+\ket{\epsilon_{g}}\bra{\epsilon_{e}}),\label{eq:H_photon_def}\\ 
L&=\sqrt{\kappa}\ket{\epsilon_{g}}\bra{\epsilon_{e}},\label{eq:L_photon_def}
\end{align}where $\Delta$ is a detuning between the laser field and the atomic-transition frequencies, $\Omega$ is the Rabi oscillation frequency, and $\kappa$ is the decay rate. We consider the counting observable $\mathcal{G}$ defined in the main text, which counts the number of photons emitted during $[0,T]$. We randomly select $\Delta$, $\Omega$, $\kappa$, $T$, and the initial density operator $\rho$. We calculate $\mathrm{Var}[\mathcal{G}]/\braket{\mathcal{G}}^2$ for the selected parameters and the density operator (the parameter ranges are shown in the caption of Fig.~\ref{fig:bound_plot}(b)). In Fig.~\ref{fig:bound_plot}(b), circles show $\mathrm{Var}[\mathcal{G}]/\braket{\mathcal{G}}^2$ as a function of $\Xi$ for many random realizations, where the dashed line denotes $1/\Xi$. Again, we see that all realizations satisfy Eq.~\mainUresult{}.

We calculate $\chi$, defined in the main text, for the two-level atom system and obtain $\chi = -\kappa \Omega \mathfrak{I}[\rho_{eg}]$, where $\rho_{eg}\equiv \braket{\epsilon_e|\rho|\epsilon_g}$ and $\mathfrak{I}[\bullet]$ returns the imaginary part of the argument. Equation~\XiUdiffUdef{} indicates that, when nondiagonal elements of $\rho$ with respect to the basis $\{\ket{\epsilon_g},\ket{\epsilon_e}\}$ do not vanish, the precision of counting observables can be increased. Note that $\chi$ can take both positive and negative values. The nondiagonal elements in the density operator are often associated with quantum coherence \cite{Streltsov:2017:CoherenceReview}, which quantifies the deviation of quantum systems from classical counterparts and has been reported to enhance the performance of thermodynamic systems, such as quantum heat engines \cite{Scully:2011:CoherentHE,Holubec:2018:QHE}. Similarly, our result shows that quantum coherence can be used to enhance the precision of counting observables for continuous measurement. In Fig.~\ref{fig:bound_plot}(c), we numerically calculate whether $\mathrm{Var}[\mathcal{G}]/\braket{\mathcal{G}}^2$ of the counting observable can be bounded from below by $1/\Xi_{\mathrm{CL}}$ and confirm that some realizations are lower than $1/\Xi_\mathrm{CL}$ (circles inside the dotted ellipse in Fig.~\ref{fig:bound_plot}(c)). This indicates that quantum coherence improves the precision of the counting observable.

\subsection{Classical Markov process}

As mentioned in the main text, Eq.~\mainUresult{} with Eq.~\XiUclassical{} holds for arbitrary time-dependent Markov processes with a transition rate $\gamma_{ji}(t)$ and initial probability $P_i$. To verify the bound, we consider the following time-dependent transition rates:
\begin{align}
\gamma_{ji}(t)=\frac{1+A\sin(\omega t+\vartheta)}{2}\nu_{ji},
\label{eq:classical_gamma_def}
\end{align}where $\nu_{ji}$ is the time-independent rate, $\omega \in \mathbb{R}$ is the angular frequency, $\vartheta \in \mathbb{R}$ is the initial phase, and $-1\le A \le 1$ is the amplitude of the oscillation. We consider the counting observable $\mathcal{G}$, as defined in the main text. We randomly select $T$, $N_S$, $\nu_{ji}$, $A$, $\omega$, and $\vartheta$ for the transition rate $\gamma_{ji}$; $P_i$ for the initial probability; and $G_{ji}$ for the counting observable (the parameter ranges are shown in the caption of Fig.~\ref{fig:bound_plot}(d)). In Fig.~\ref{fig:bound_plot}(d), the circles show $\mathrm{Var}[\mathcal{G}]/\braket{\mathcal{G}}^2$ as a function of $\Xi_\mathrm{CL}$ for many random realizations, where the dashed line denotes the lower bound $1/\Xi_\mathrm{CL}$. Here, $\Xi_\mathrm{CL}$ is defined in Eq.~\XiUclassical{}. 
It is shown that all realizations are above the lower bound, verifying that the main result holds for classical time-dependent Markov processes.

When the system is in a steady state, it is known that the fluctuations in the counting observable are bounded from below by $1/\Upsilon$, where $\Upsilon$ is the dynamical activity \cite{Garrahan:2017:TUR}. Thus, we also check whether $\mathrm{Var}[\mathcal{G}]/\braket{\mathcal{G}}^2$ can be bounded by the dynamical activity for the time-dependent case. In Fig.~\ref{fig:bound_plot}(e), we plot $\mathrm{Var}[\mathcal{G}]/\braket{\mathcal{G}}^2$ as a function of $\Upsilon$ using circles and $1/\Upsilon$ using the dashed line. We observe that some realizations are below $1/\Upsilon$ (circles inside the dotted ellipse in Fig.~\ref{fig:bound_plot}(e)). To highlight this violation, we plot $\Upsilon\mathrm{Var}[\mathcal{G}]/\braket{\mathcal{G}}^2$ as a function of $\Upsilon$ with  circles in Fig.~\ref{fig:bound_plot}(f), where the dashed line describes $1$. When $1/\Upsilon$ is the lower bound of $\mathrm{Var}[\mathcal{G}]/\braket{\mathcal{G}}^2$, all realizations should be above the dashed line in Fig.~\ref{fig:bound_plot}(f). Some realizations are clearly below 1, indicating that $\mathrm{Var}[\mathcal{G}]/\braket{\mathcal{G}}^2$ cannot be bounded from below by $1/\Upsilon$ for time-dependent Markov processes.